\documentclass[journal]{IEEEtran}
\usepackage{graphicx}
\usepackage{url}
\usepackage{amsmath, amsfonts}
\usepackage{amsthm, amssymb}
\usepackage{comment}
\usepackage{xcolor}
\usepackage[font=small]{caption}
\usepackage{subcaption}
\usepackage{array}
\usepackage{bbm}
\usepackage{cite}

\usepackage{pifont}

\DeclareMathAlphabet\mathbfcal{OMS}{cmsy}{b}{n}

\ifCLASSINFOpdf
\else
\fi
\begin{document}

\title{Multi-Modal Sensing and Fusion in mmWave Beamforming for Connected Vehicles: \\ A Transformer Based Framework}

\author{Muhammad Baqer Mollah,~\IEEEmembership{Graduate Student Member,~IEEE,}
        Honggang Wang,~\IEEEmembership{Fellow,~IEEE,}
        \\ Mohammad Ataul Karim,~\IEEEmembership{Life Fellow,~IEEE,}
        and Hua Fang,~\IEEEmembership{Senior Member,~IEEE}%

\thanks{Muhammad Baqer Mollah and Mohammad Ataul Karim are with the Department of Electrical and Computer Engineering, University of Massachusetts Dartmouth, MA 02747, USA (Emails: mmollah@umassd.edu; mkarim@umassd.edu).}
\thanks{Honggang Wang and Hua Fang are with the Department of Graduate Computer Science and Engineering, Katz School of Science and Health, Yeshiva University, NY 10016, USA (Emails: honggang.wang@yu.edu; hua.fang@yu.edu). (Corresponding Author: Honggang Wang)}
\thanks{The preliminary version of this work has been accepted and presented at IEEE Global Communications Conference (IEEE GLOBECOM) 2025 \cite{mollah2025transformers}.}
\thanks{This research is supported in part by National Science Foundation (NSF) under the grants 2010366 and 2140729.}
\thanks{This article has been accepted for publication in IEEE Transactions on Vehicular Technology. This is the author's version which has not been fully edited and content may change prior to final publication.}
\thanks{Digital Object Identifier: 10.1109/TVT.2026.3665294}
\thanks{\copyright 2026 IEEE. Personal use of this material is permitted. However, permission to use this material for any other purposes must be obtained from the IEEE by sending a request to pubs-permissions@ieee.org.}%
}
\markboth{IEEE Transactions on Vehicular Technology}%
{Shell \MakeLowercase{\textit{et al.}}: Bare Demo of IEEEtran.cls for IEEE Journals}

\maketitle

\begin{abstract}
    Millimeter wave (mmWave) communication, utilizing beamforming techniques to address the inherent path loss limitation, is considered as one of the key technologies to support ever increasing high throughput and low latency demands of connected vehicles. However, adopting standard defined beamforming approach in highly dynamic vehicular environments often incurs high beam training overheads and reduction in the available airtime for communications, which is mainly due to exchanging pilot signals and exhaustive beam measurements. To this end, we present a multi-modal sensing and fusion learning framework as a potential alternative solution to reduce such overheads. In this framework, we first extract the representative features from the sensing modalities by modality specific encoders, then, utilize multi-head cross-modal attention to learn dependencies and correlations between different modalities, and subsequently fuse the multimodal features to obtain predicted top-$k$ beams so that the best line-of-sight links can be proactively established. To show the generalizability of the proposed framework, we perform a comprehensive experiment in four different vehicle-to-infrastructure (V2I) and vehicle-to-vehicle (V2V) scenarios from real world multimodal and 60 GHz mmWave wireless sensing data. The experiment reveals that the proposed framework (i) achieves up to $96.72$\% accuracy on predicting top-15 beams correctly, (ii) incurs roughly $0.77$~dB average power loss, and (iii) improves the overall latency and beam searching space overheads by $86.81$\% and $76.56$\% respectively for top-$15$ beams compared to standard defined approach.
\end{abstract}

\begin{IEEEkeywords}
    Beamforming, connected vehicles, transformers, attention mechanism, millimeter-wave communications, multi-modal fusion.
\end{IEEEkeywords}

\vspace{-10pt}

\IEEEpeerreviewmaketitle

\section{Introduction}
    \IEEEPARstart{F}{uture} transportation systems will be revolutionized by connected and autonomous vehicles capable of greatly improved traffic efficiency, mobility, and transportation safety, where seamless connectivity with high throughput and lower latency is crucial for extensive sensor data collection and sharing among vehicles, infrastructures, and surroundings \cite{cheng2025driving}. Such sensor data sharing plays an important role in safety critical tasks to support self-driving features such as high-definition map updating for safer maneuvering of the vehicles and cooperative perception for managing blind spots. For example, according to the report in \cite{misc1}, an approximate $25$ gigabytes/hour is required by a connected vehicle for the purpose of sensor data communication, and which may approach up to $500$ gigabytes/hour once fully connected and autonomous.

    The 5G New Radio (5G-NR) standard has been specified to operate in both sub-6 GHz and higher frequencies ($> 24$ GHz) at millimeter-wave (mmWave) band, however, the sub-6 GHz band is too congested already for use in these instances. Rather, the mmWave band has been shown as an ideal candidate due to having an abundant spectrum resources toward fulfilling the high throughput and lower latency demands \cite{mollah2024mmwave,yu2024optimization}. In general, signals in mmWave band suffer from severe atmospheric path loss due to having short wavelengths (roughly 1$\sim$10 mm) \cite{osa2023measurement}. For this reason, beamforming techniques are utilized in mmWave systems by deploying phased-antenna arrays and configuring a narrow beam of RF energy to establish and maintain high directional links, thereby ensuring sufficient signal strength.
        
    The 5G-NR standard, on the other hand, defines a beam training approach based on codebooks over phased-antenna arrays \cite{xue2024survey,ullah2025birc}. Typically, the codebook-based approach enables the communicating nodes to coordinate each other by exchanging pilot signals and measurements for selecting the best beam direction. However, when it comes to vehicular moving scenarios where the beam directions change dynamically, this process often introduces highly undesirable computational and latency overheads. As an example, according to \cite{salehi2022deep}, it may take around $25$ milliseconds to find out a suitable beam pair among $60$ beams in the 5G-NR standard based mmWave system and requires it to be repeated when the vehicles move forward. Besides, the overheads may also increase linearly with the number of beam directions in codebooks. Accordingly, it becomes essential to save such beam training overheads in order to fully unlock the potential benefits from mmWave communications in connected vehicles domain.
        
    Given the aforementioned challenge in the context of vehicular settings, recent works have focused on utilizing out-of-band side information as an alternative promising approach to configure the mmWave communications links. As such, this side-information can be extracted from non-RF sensing sources, such as position \cite{xiu2025robust,mollah2024position}, visual \cite{charan2025camera,zhang2025rfsoc}, and Light Detection and Ranging (LiDAR) \cite{ohta2025real2sim2real,zhang2025synesthesia} sensors, however, each of these single-modal sources has its own limitations, which essentially makes them often impractical to use individually. In contrast, utilizing more than one sensing sources, that is multi-modal sensing, offers more robust and generalization capabilities by taking the full advantages from each modality, and according to \cite{bai2025multi,wang2025multi,luo2025isac} several approaches have been introduced based on multi-modality sensing in recent years. However, most of these works are either limited to considering relatively simpler V2I scenarios instead of dynamic V2V scenarios or validating by ray-tracing simulations. Most recently in \cite{mollah2025multi}, a unified multi-modal deep learning based solution for both V2I and V2V communications has been demonstrated, and subsequently validated with real-world data.
    
\subsection{Transformers on Multi-Modal Learning}
    Drawing inspiration from the recent successes, the transformer architectures, initially introduced in \cite{vaswani2017attention} followed then by many advanced variants, have widely been adopted in various fields due to its renowned attention mechanism, with foundation models, edge intelligence, wireless sensing, wireless resource allocation, vision, point clouds, and multimodal learning being few notable examples among others. As such, several solutions have incorporated transformer-based techniques on multi-modal learning to facilitate mmWave beamforming. For example, Cui et al. \cite{cui2024sensing} proposed a multimodal fusion among the modality specific features by employing transformer's merged-attention technique before performing final processing by the multi-layer perceptron. Likewise, the solution in \cite{tariq2024deep} utilized quantum-transformer networks including quantum embeddings and classical deep learning techniques to extract from the multi-modal sensory data, and then performed a straightforward concatenation for multimodal features fusion on the resultant outputs from the quantum, convolutions, and transformer modules to obtain the estimated beam indices. Meanwhile, Luo et al. \cite{luo2025vision} have introduced a beam selection approach by fusing visual and LiDAR sensing information, where the key idea is to utilize multi-modal group attention transformer, an attention variant that processes the relationships between multiple entities within a group, to merge the intermediate representations.
    
    Further, the authors of \cite{raha2025advancing} have proposed a multi-modal solution by considering the transformer architecture along with semantic localization technique, however, this work has particularly focused on improving the robustness under different weathers, such as rainy, foggy, sunny, and snowy conditions. Besides, Ghassemi et al. \cite{ghassemi2024multi} have emphasized on a robust beam management approach having improved adaptability in dynamic environments. For this approach, the authors introduced a framework by integrating multi-modal transformers based encoders with a deep Q-learning technique design, where the available beams are divided into multiple groups first and then utilize multi-modal transformers to determine the optimal group, and finally, the deep Q-learning helps to make fast and adaptable decision on beam tracking, thereby ensuring maximized throughput. Similar objective has been considered in \cite{farzanullah2025beam}, however, for beam selection in Integrated Sensing and Communication (ISAC) scenario using a multi-agent contextual bandit algorithm, a reinforcement learning based approach that makes informed decisions from the contextual information and interactions by multiple agents. In particular, this approach adopted three steps: first utilizes transformers based encoders to learn the inter-modality interactions between user location and ISAC sensing information; then, uses the multi-agent contextual bandit algorithm to learn the environment for single user setting; and finally utilizes transfer learning to reduce the training time for multi-user settings. Park et al. \cite{park2025resource} have proposed a resource-efficient beam prediction approach by knowledge distillation concept, that is, transferring the knowledge from a transformers based multi-modal teacher model to a lightweight uni-modal model as student. The simulation on CARLA platform and generated synthetic mmWave channels results showed that the student model can obtain roughly $94.62$\% of the prediction performance of teacher with just $10$\% of its model parameters, which in turn contributes to significant saving in computational resources.

\subsection{Motivations and Contributions}
    Nevertheless, these prior solutions on beamforming tasks typically tend to employ fusion at the feature representation level while applying interactions only subsequent to the pooling operation and seldom consider inter-modality relationships, i.e., the correlations between different sensing modality features. Such fusion techniques thus inherently overlook a considerable detailed information necessary for optimal performance, thereby compromising the performances on prediction and robustness. And with this in mind, we introduce a multi-modal learning framework for pursuing proactive beam selection solution for mmWave enabled V2I and V2V communications. The main contributions of this work are summarized as follows:
    
    \begin{itemize}
    \item Motivated by the advantages of utilizing non-RF sensing modalities, a multi-modal sensing and fusion framework for mmWave beam selection is designed that predicts the top-$k$ beams, that is, a subset of total beam on codebooks, which helps to considerably reduce the beam searching space and end-to-end latency.
    
    \item To overcome the limitations of individual sensing capabilities in practical scenarios, multiple sensing information that can be obtained from position, visual, and LiDAR sensors, commonly available on road-side units and modern vehicles is employed, thereby reducing the complicacy and providing robustness in terms of missing information and errors.
    
    \item Unlike straightforward fusion of multimodal features introduced in other solutions, richer representations are achieved by employing multi-head cross-modal attention between the features of visual and LiDAR modalities, where the cross-attention in particular facilitates to learn the dependencies and correlations between them.
    
    \item The effectiveness of the proposed framework is evaluated on four diverse V2I and V2V real world scenarios. The results show that the framework performs favorably better than state-of-the-art baseline approaches in terms of prediction accuracies and averages power losses, hence implying the necessity of utilizing multi-head cross-modal attention.
    
    \item The proposed framework shows that it can be potentially integrated with 5G-NR standard contributing in turn to reduced beam searching and end- to-end latency overheads.
    \end{itemize}

\subsection{Organization and Notations}
    The rest of this paper is organized as follows: Section II describes the considered system model and describing the beam selection problem. Section III provides the design of the proposed framework including sensing modalities, solutions methodology, and preprocessing of sensing modalities. Section IV details the constructions that leverages modality specific encoders, fusion by multi-head cross-modal attention, and beam selection modules. Section V presents the experimental results. Section VI discusses the integration with 5G-NR standard. Finally, Section VII concludes the paper. The Table~\ref{tab: notations} summarizes the list of major notations used in this paper.

\section{System Model and Problem Description}
    In this section, we present the considered system model including mmWave enabled vehicular communications, followed by a discussion of the beam selection problem by multi-modal sensing addressed in this work.
    
\subsection{System Model}
\begin{figure} [!t]
    \includegraphics[width=\linewidth]{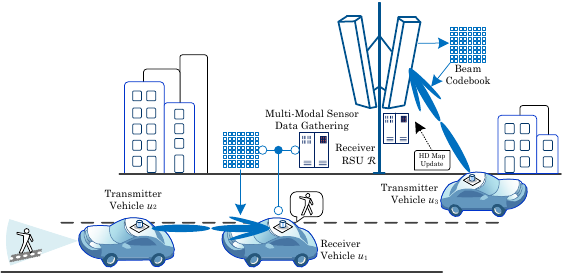}
    \caption{Illustration of our considered system model enabled by mmWave V2I and V2V communications.}
    \label{fig: system}
\end{figure}

    Fig.~\ref{fig: system} illustrates a downlink vehicular communication system with three moving vehicles (one receiving and two transmitting) characterized respectively with $u_1$, $u_2$, $u_3$ and a road-side unit (RSU) $\mathcal{R}$. Herein, the primary receiving moving vehicle $u_1$ is connected by vehicle-to-vehicle (V2V) communication link to the transmitting vehicle $u_2$, whereas the RSU $\mathcal{R}$ is connected to the passing transmitting vehicle $u_3$ by vehicle-to-infrastructure (V2I) communication link to enable communication services, such as cooperative perceptions and high-definition map data crowdsourcing withing their respective coverage areas. Specifically, the units $u_1$ and $\mathcal{R}$ are acting as receivers, while the units $u_2$ and $u_3$ are acting as transmitters. In order to capture multi-modal sensing information from the surrounding environment, the receiver units are equipped with camera and LiDAR sensors, whereas the transmitter units carry the GPS receivers.
    
    Here, $u_1$ and $\mathcal{R}$ are assumed to be equipped with (an) phased arrays employing $N_{rx}$ antenna elements uniform linear arrays (ULA) operating at mmWave frequency bands. Similarly, the other two units $u_2$ and $u_3$ are equipped with $N_{tx}$ elements antennas. For the purpose of simplicity, $N_{tx} \gg N_{rx}$ is assumed, and subsequently $N_{tx}$ is set to $1$, which refers the omnidirectional transmissions realized using the single antenna element of the phased array. However, each phased arrays at receiver sides employs a pre-defined over-sampled beamsteering codebooks $\mathbfcal{C}_{rx} = \{\mathsf{c}_1, \mathsf{c}_2, ...., \mathsf{c}_K\}$ containing $K$ number of total beam directions toward receiving the transmitted signals. Here, $\mathsf{c}_{\kappa} \in \mathbb{C}^{N_{rx} \times 1}$ represents any beamsteering vector, which can be expressed as
\begin{equation}
    \mathsf{c}_{\kappa} = \frac{1}{\sqrt{N_{rx}}} \biggr[1, e^{j\frac{2\pi}{\lambda}d\sin(\phi_{\kappa})}, ...., e^{j(N_{rx} - 1)\frac{2\pi}{\lambda}d\cos(\phi_{\kappa})}\biggr]^T,
\end{equation}
    where, $\lambda$, $d$, and $\phi_{\kappa}$ represent the carrier wavelength, the antenna spacing, and the azimuthal angles uniformly quantized within $[-\pi/4, \pi/4]$ range, respectively.
    
    Further, we consider the downlink communication systems adopting orthogonal frequency-division multiplexing signal transmissions model. Let, $\mathcal{Q} = \{1, ...., q_m\}$ represents the set of subcarrier indices with $q_m$ total number of subcarriers in a time-varying channel, $s \in \mathbb{C}$ is the transmitted data symbol, $\mathbb{E}|s|^2$ = $P_s$ (average power), and $\mathbf{h} \in \mathbb{C}^{N_{rx} \times 1}$ indicates the downlink channels of the V2V and V2I links. Considering the 3D geometric channel model in this work, the $\mathbf{h}$ can be written in the frequency domain as 
\begin{equation}
    \mathbf{h} = \sqrt{N_{rx}} \sum_{d=1}^{D}\sum_{l=1}^{L}\alpha_l e^{-j\frac{2\pi q}{\mathcal{Q}}d} p(dT_s - \tau_l)\textbf{a}_{rx} (\theta_{l}^A, \phi_{l}^A).
\end{equation}
    
    Here, $\sqrt{N_{rx}}$, $L$, $C$, $p(\cdot)$, $T_s$, and $\textbf{a}_{rx}(\cdot)$ represent the normalization factor, the number of total available propagation paths, the cyclic prefix length, the pulse shaping filter, the symbol period, and the function of receiver's complex beam steering vectors, respectively while $\alpha_l$, $\tau_l$, $\theta_{l}^A$, and $\phi_{l}^A$ are the complex path gain, the delay, elevation angle of arrival, azimuth angle of arrival, respectively, of each channel path $l$. 
    
\subsection{Problem Description}
    Given the considered system model, the receiving vehicle $u_1$ and RSU $\mathcal{R}$ observe the corresponding downlink received signal at time $t$ and $q^{\text{th}}$ subcarrier as
\begin{equation}
    r_t[q] = \mathbf{h}_t^\mathsf{T}[q] \mathsf{c}_{\kappa} s_t[q] + n_t[q], \quad \forall q \in \mathcal{Q},
\end{equation}
    where, $(\cdot)^\mathsf{T}$ is the conjugate transpose, and the variable $n \sim \mathcal{C}\mathcal{N}(0, \sigma_n^{2})$ denotes the received complex Gaussian noise sample with zero mean and $\sigma_n^{2}$ variance. With the fixed codebook constraint, let $\mathcal{B}_{rx} = \{\mathsf{b}_1, \mathsf{b}_2, ...., \mathsf{b}_K\}$ with $|\mathcal{B}_{rx}| = K$ is the set of all possible beams to perform beamforming by the vehicle $u_1$ and RSU $\mathcal{R}$ to obtain adequate received powers, and $\mathcal{P}_{rx} = \{\mathsf{p}_1, \mathsf{p}_2, ...., \mathsf{p}_K\}$ is the set of corresponding normalized received power, where the received power for a specific beam $\mathbf{p}_{\kappa}$ can be calculated by summing over all subcarriers. Then, the beam selection task is to find the optimal beam $\mathsf{b}_{\kappa}^*$ out of $\mathcal{B}_{rx}$ that maximizes the received power. Mathematically, this beam selection problem can be formulated as follows:
\begin{equation}
    \mathsf{b}_{\kappa}^* = \underset{\mathsf{c}_{\kappa} \in \mathcal{C}_{rx}}{\arg\max} \sum_{q=0}^{\mathcal{\mathcal{Q}}-1} |\mathbf{h}_t^\mathsf{T}[q] \mathbf{c}_{\kappa} s_t[q]|^2.
\end{equation}
    
    In 5G-NR standard defined beam selection approach, finding the optimal beam $\mathsf{b}_{\kappa}^*$ requires sweeping all beams sequentially over the beam codebook, which typically introduces high beam training and latency overheads. However, the aim of this work is to investigate how to utilize out-of-band sensing information to address this problem by predicting a set of recommended beams $\mathcal{B}_{k} = \{\mathsf{b}_i\}_{i=1}^k$ as top-$k$ beams such that $\mathcal{B}_{k} \subset \mathcal{B}_{rx}$ and $\mathsf{b}_{\kappa}^* \in \mathcal{B}_{k}$ as top-1 beam.
    
\section{Design of the Proposed Framework}
    In this section, we describe the design overview of the proposed transformer based beam selection framework by multi-modal sensing and fusion in details, highlighting the key steps, such as sensing modalities, solution methodology, and finally preprocessing of sensing modalities.
    
\subsection{Multimodal Sensing Modalities}
    The proposed framework leverages multimodal sensing modalities, which are captured from the following four sensing elements:

    \subsubsection{Position Sensors}
    The Position sensors provide precise real-time locations of the vehicles by Global Positioning System (GPS) receiver. The readings from such position sensors include latitude and longitude representations in decimal degrees with the positive values for north and east and the negative values for south and west in geographic coordinates. This sensory information is considered as a key component of this work, enabling them to make targeted vehicular user identification.
    
    \subsubsection{Visual Sensors}
    The visual sensors, such as RGB/RGB-depth cameras, are the main visual perception elements. These sensors generate continuous frames in the form of visual snapshots from the surroundings of road-side units and vehicles to provide a comprehensive environmental assessment. Typically, the field of view of such cameras range from 110 to 360-degree.
    
    \subsubsection{LiDAR Sensors} 
    The LiDAR sensors offer detailed perception of the environment, which is necessary to understand the the dynamic road conditions. Such sensors produce three-dimensional spatial data with detailed measurements of distances from the sensor location to the surrounding objects. The distance measurements is obtained by emitting pulsed lasers and calculating the reflecting times. At discrete time intervals, the outputs are represented as 3D point clouds, a vast collection of points.
    
    \subsubsection{Wireless Sensors} 
    The mmWave phased arrays are utilized as the wireless sensors, where each phased array consists of a set of small antennas placed next to each other as an array with the over-sampled and pre-defined codebook of beam directions. Then, we can get the outputs from these sensors are the measured received powers in each of the beam directions under the codebook.

    Finally, we define the multi-modal samples captured by the aforementioned sensors at timestamp $t$ as $X_{g}[t] \in \mathbb{R}^{2}$ (latitude and longitude), $X_{v}[t] \in \mathbb{R}^{w_v \times h_v \times c_v}$ (width, height, and number of color channels), $X_{p}[t] \in \mathbb{R}^{d_p \times h_p \times w_p}$ (depth, height, and width), and $Y_i$, respectively, where $Y_i$ in particular represents the optimal beam label across the codebook. During the data acquisition phase, the road-side unit $\mathcal{R}$ and connected vehicle $u_1$ are required to collect those samples in different scenarios and subsequently build a dataset, $\mathcal{D} = \{(X_i, Y_i)\}_{i=1}^N$, where $X_i = (X_{g}, X_{v}, X_{p})$ and $N$ is the number of total samples.

\begin{table}[!t]
    \centering
    \caption{Notation Summery.}
    \label{tab: notations}
    \begin{tabular}{m{1.4cm}|m{5.7cm}}
    \hline \hline
    \textbf{Notation}   &   \textbf{Description} \\
    \hline

    $u_1$, $u_2$, $u_3$    &    Moving vehicles \\
    \hline

    $\mathcal{R}$    &    Road-side unit, gNodeB, or base station \\
    \hline
     
    $N_{rx}$     &    Antenna elements in phased arrays at receiver end \\
    \hline

    $\mathbfcal{C}_{rx}$     &    Beamsteering codebooks \\
    \hline
    
    $K$     &    Number of total beam directions \\
    \hline

    $\mathcal{Q}$    &    The set of subcarrier indices \\
    \hline

    $\mathbf{h}$    &    Downlink channel matrix of communication links \\
    \hline

    $r_t$    &     Downlink received signal at time $t$ \\
    \hline
    
    $\mathcal{P}_{rx}$    &    The set of normalized received powers \\
    \hline
    
    $X_{g}, X_{v}, X_{p}$    &    Multi-modal samples captured by the sensors \\
    \hline

    $\mathbb{X}_{g}^{n}, \mathbb{X}_{v}^{n}, \mathbb{X}_{p}^{n}$    &    New sensing modality inputs for inference/testing \\
    \hline

    $Y_i$    &    The optimal beam label across the codebook \\
    \hline

    $\omega$    &    Learnable parameters \\
    \hline
    
    $\mathcal{M}(\cdot, \omega)$    &    The multimodal model \\
    \hline
    
    $\mathcal{M}_{{\omega}^*}$    &    Trained model with optimal parameters $\omega^*$ \\
    \hline
    
    $\hat{\mathsf{b}}$    &    Probabilistic predictions of top-$k$ beams \\
    \hline
    
    $T_{sp}$    &    Beam sweeping time in 5G-NR \\
    \hline
    
    $T_{sp}^{mm}$    &    Sweeping time for the top-$k$ predicted beams \\
    \hline
    
    $\mathcal{T}$     &    Total time required to complete the beam selection \\
    \hline
    \end{tabular}
\end{table}

\subsection{Solution Methodology}
    Once the dataset $\mathcal{D} = \{(X_i, Y_i)\}_{i=1}^N$ has been built, the beam selection task is solved as defined in Section II-B by the proposed multi-modal sensing assisted framework. For that, we set up a multimodal model $\mathcal{M}(\cdot, \omega)$ to learn the mapping of the multi-modal sensing inputs (position, visual, and point clouds) to the top-$k$ beams, where $\omega$ is the learnable parameters which can be learned (i.e., adjusted) by training. Toward this objective, the multimodal model can be optimized to find the optimal model parameter $\omega^*$ with the aim of minimizing the loss between the model output and optimal beam labels (ground truth beams). In this work, we train the model by supervised learning so that that the useful features can be extracted from $X_i$ with generalization abilities and the optimal model parameters can be learned from all training samples of the dataset. Mathematically, the optimization problem can be expressed as follows:
\begin{equation}
    \omega^* = \underset{\omega}{\arg\min} \frac{1}{|\mathcal{D}|}\sum_{(X_i, Y_i) \in \mathcal{D}} \mathcal{J}(\mathcal{M}(X_i, \omega), Y_i),
\end{equation}
    where, $\mathcal{J}(\cdot)$ represents the loss function which measures the difference between the predicted (estimated) beams from the model and the ground truth beams. Finally, to convert the output of the model $\mathcal{M}(X_i, \omega)$ into a probabilistic distribution form (ordered as the largest to the smallest), a softmax function $\mathsf{Softmax}$ is applied on it to obtain the predicted results as $\hat{\mathsf{b}} = \mathsf{Softmax}(\mathcal{M}(X_i, \omega))$. However, to solve the problem defined in Eq. (5), cross entropy as the loss function is utilized during the training phase given by
\begin{equation}    
    \mathcal{J}(\hat{\mathsf{b}}, Y) = - \frac{1}{N}\sum_{i=0}^{N-1}\sum_{k=0}^{K-1} Y_{ik}\log(\hat{\mathsf{b}}_{ik}),
\end{equation}
   where $K$ and $\hat{\mathsf{b}}$ represent the total number of beams and the predicted output from the model, respectively. After the model training process, the most up-to-date trained model can be utilized to infer beams by the vehicle $u_1$ and RSU $\mathcal{R}$ to serve their respective connected vehicle $u_2$ and passing vehicle $u_3$ within their coverage areas. An important point is that the trained model should have the generalization ability to work on new sensing modality inputs, such as GPS coordinate, visual, and point cloud samples $\mathbb{X}_{g}^{n}$, $\mathbb{X}_{g}^{n}$, and $\mathbb{X}_{lid}^{n}$, respectively while deploying in practical environment. With these new inputs, the probabilistic predictions of top-$k$ beams can be obtained as:
\begin{equation}
    \hat{\mathsf{b}} = \mathsf{Softmax}(\mathcal{M}_{{\omega}^*}(\mathbb{X}_{g}^{n}, \mathbb{X}_{v}^{n}, \mathbb{X}_{p}^{n})),
\end{equation}
    where, $\mathcal{M}_{{\omega}^*}$ is the trained model with optimal parameters $\omega^*$. However, in practical environment, certain prediction errors may be introduced due to dynamic nature of the surroundings. Hence, for addressing such performance degradation, the model might be updated with relatively smaller dataset compared to the used dataset for initial training and fine-tuned according to the performance requirements.
       
\subsection{Preprocessing of Sensing Modalities}
    The collected raw modality samples, specifically location coordinates, visual, and point clouds are desired to be compatible to the model and properly formatted while being used as inputs to the model. Hence, before feeding these raw samples into the subsequent proposed multimodal model (details are presented in next section), proper preprocessing procedures are necessary for both training and inference steps. In particular, the preprocessed samples can facilitate on speeding up and stabilized learning during the model training. We now describe the preprocessing steps considered in this work as follows.

    \subsubsection{Coordinate Samples}
    The coordinate samples are basically in two-dimensional space, where each point is represented by latitude and longitude, and their range of values are within $-90$ to $+90$-degree $-180$ to $+180$-degree, respectively. For the purpose preprocessing of these samples, we apply a normalization technique across the latitude and longitude values with a min-max scaling. To get normalized values of latitude $g_{lt}^{\prime}$ and longitude $g_{lg}^{\prime}$ at time $t$, this min-max scaling technique first subtracts the minimum value from each data point value, and then dividing the result by the range (the difference between maximum and minimum values), which can be defined as: $(g_{lt}^{\prime}[t], g_{lg}^{\prime}[t]) = ((g_{lt} - min)/(max - min), (g_{lg} - min)/(max - min)$, where $g_{lt}$ and $g_{lg}$ are the original latitude and longitude values, respectively.
    
    \subsubsection{Visual Samples}
    Depending on the visual sensors and their purposes, the captured visual image samples may have typical resolutions of $1280 \times 720$, $1920 \times 1080$, $2560 \times 1440$, or $3840 \times 2160$. However, the images with higher resolutions contains more pixels, which require increasing computational processing time and memory usages. Hence, we first resize the sample images into a reduced and same sized spatial dimensions of $224 \times 224$, where $224 \leq w_v$ and $224 \leq h_v$, then the total numbers of pixels become $224 \times 224 = 49,536$. After that, we normalize the red, green, and blue color channel intensities (each range from 0-255) to have zero mean and unit variance by subtracting a fixed mean $(0.485, 0.456, 0.406)$ and dividing the results by a fixed standard deviation $(0.229, 0.224, 0.225)$ for each color channels.
    
    \subsubsection{Point Cloud Samples}
    From the data structure perspective, a point cloud sample is a set of data points in 3D space. As a result of measurement for each scan, the LiDAR sensor produces a vast amount of points $P = \{(x_p, y_p, z_p)\}_{p=1}^{\mathsf{n}}$, where each point $\{(x_p, y_p, z_p)\}$ represents the coordinate of an obstacle point calculated from the reflection of emitted laser pulses. However, the raw point cloud samples are data intensive, and the number of points in each sample varies according to the environment. For example, highways with fewer reflections produce fewer points while urban areas having complex surfaces, buildings, and vehicles introduce higher density of points. Therefore, we preprocess the raw sample $P$ to obtain a desired point density by either padding or downsampling, and accordingly we set $\mathsf{n} = 15,000$ fixed number of points. Depending on the point numbers, if any sample has lesser points than the desired points, we pad the rest of points with zeros, whereas we downsample by eliminating random points otherwise.

    To this end, we can express the aforementioned preprocessed multi-modal samples at timestamp $t$ as $X^{\prime}_{g}[t]$, $X^{\prime}_{v}[t]$, and $X^{\prime}_{p}[t]$, respectively. With these preprocessed samples, we briefly explain the construction of the proposed framework in the following section.

\section{Detailed Construction}
\begin{figure*} [!t]
	\includegraphics[width=.95\linewidth]{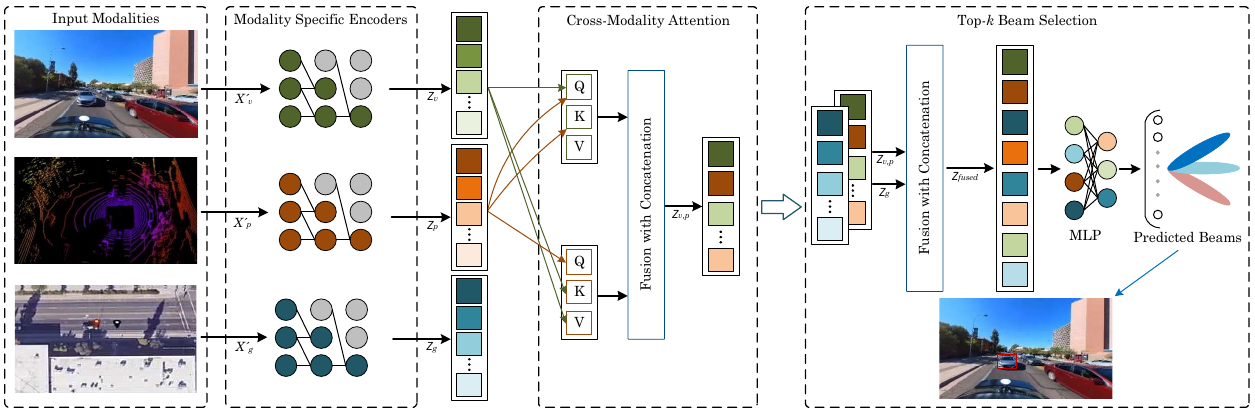}
    \centering
    \caption{The overview of the proposed framework, which is comprised of three distinct modality specific encoders for feature extractions, each taking the pre-processed modalities as inputs, followed by multi-modal fusion and beam prediction procedures.}
    \label{fig: transformer-Model}
\end{figure*}

    Fig. \ref{fig: transformer-Model} illustrates the overall framework, which is comprised of three input modalities; each pre-processed modalities are processed next using distinct modality specific encoders for feature extractions. The extracted features are fed next into a multi-head cross-modality attention module (instead of multi-head, single attention shown for simplicity). The extracted features are subsequently passed through the top-$k$ beam selection module for final predictions. The specific details of these modules are presented in the following subsections.
    
\subsection{Modality Specific Encoders}
    Before proceeding to multimodal fusion, the features are extracted independently by employing the following modality specific encoders with a focus on extracting the most relevant features from each modality.
    
    \textit{1) Position Encoder:} For constructing the position encoder, two components are adopted from the transformer architecture: an embedding layer and a transformer encoder, customized for GPS coordinates feature extraction modeling. To be specific, given the preprocessed position data $X^{\prime}_{g} \in \mathbb{R}^2$, the embedding layer first transforms each input feature from $X^{\prime}_{g}$ into a higher-dimensional feature space (i.e., $d_g = 512$) in order to enable richer feature learning, which can be expressed as
\begin{equation}
    \mathbf{E}_g = X^{\prime}_{g}\mathbf{W}_{g} + \mathbf{b}_{g},
\end{equation}
    where, $\mathbf{W}_g \in \mathbb{R}^{d_g \times 2}$ and $\mathbf{b}_{g}$ are the weight matrix and bias, respectively. After obtaining the higher-dimensional features, a multi-layered transformer encoder equipped with $6$ stacked layers, each layer composed of a multi-head attention sub-layer (each with $h$ = $8$ number of attention heads) is then utilized, followed by a feed-forward sub-layer. Particularly for the calculation of attention score for single attention, the respective query ($\mathbf{Q}_g$), key ($\mathbf{K}_g$), and value ($\mathbf{V}_g$) matrices are derived from the output of embedding layer as follows:
\begin{equation}
    \mathbf{Q}_g = \mathbf{E}_g\mathbf{W}_{g}^{Q}, \quad
    \mathbf{K}_g = \mathbf{E}_g\mathbf{W}_{g}^{K}, \quad
    \mathbf{V}_g = \mathbf{E}_g\mathbf{W}_{g}^{V}.
\end{equation}
    Here, $\mathbf{W}_{g}^{Q}$, $\mathbf{W}_{g}^{K}$, and $\mathbf{W}_{g}^{V}$ are the learnable parameters during training, whereas the dimensionality of both $\mathbf{Q}_g$ and $\mathbf{K}_g$ is $d$ and $\mathbf{V}_g$ is $d_v$. Subsequently, the attention score is obtained by computing the following scaled dot-product self-attention operation equation:
\begin{equation}
    \mathsf{Attention}(\mathbf{Q}_g, \mathbf{K}_g, \mathbf{V}_g) = \mathsf{Softmax}\left(\frac{\mathbf{Q}_g\mathbf{K}_g^{\mathsf{T}}}{\sqrt{d}} \right)\mathbf{V}_g,
\end{equation}
    where, the division by $\sqrt{d}$ is used for scaling to avoid the attention scores from becoming excessively small or high, thereby enabling better convergence during training.
    
    However, aiming to capture richer contextual relationships by focusing on the different parts of the input at once, the attention operation (or head) are calculated $h$ times in parallel within the multi-head attention. This can be obtained by linearly transforming (also referred as projection) the $\mathbf{Q}_g$, $\mathbf{K}_g$, and $\mathbf{V}_g$ with $h$ times by using different learnable matrices. After that, considering the dimensions $d = d_v = d_g/h = 64$ for each of these, the results from all heads are concatenated and projected once again as following equation:
\begin{equation}
    \mathsf{MHA}(\mathbf{Q}_g, \mathbf{K}_g, \mathbf{V}_g) = \mathsf{Concat}(head_1,...,head_h)\mathbf{W}^o,
\end{equation}
    where, $head_l = \mathsf{Attention}(\mathbf{Q}_{gl}, \mathbf{K}_{gl}, \mathbf{V}_{gl})$ is the $l$-th head and $\mathbf{W}_o \in \mathbb{R}^{hd_v \times d_g}$ is the parameter matrix used for the projection. Finally, the multi-head attention results are passed through the feedforward network sub-layer to improve the feature expressiveness, thereby producing the desired features, denoting as 
    $\mathsf{Z}_{g}$.
    
    \textit{2) Visual Encoder:} To implement the visual encoder, we employ a vision transformer variant, named multi-axis vision transformer (MaxViT) model \cite{tu2022maxvit} and fine-tune according to the task. This model introduces a unified design by taking benefits from both convolution and transformer to extract rich visual features (representations) while better adapting capabilities to high-resolution and dense prediction tasks. Particularly, the MaxViT model comprises of three components, such as convolution, mobile inverted bottleneck convolution (MBConv), and multi-axis attentions. With these components, the model takes $X^{\prime}_{v}$ as input and processes it through the following ways.
    
    At first stage, the convolutional operations with $3 \times 3$ kernels are performed on the input $X^{\prime}_{vis}$ for initial feature extractions to capture simple basic features, and also downsampled with a stride of $2$, that is, reducing the heights and widths dimensions by a factor of $2$, at the same time, increasing the number of channels for subsequent complex and abstract feature extractions.
    
    The resulting outputs are then passed through a series of MaxViT blocks across four stages, where each MaxViT block incorporates the MBConv and multi-axis attentions. Here, the MBConv uses depthwise separable convolution with an additional squeeze and excitation, and follows inverter residual block structure. With depthwise and pointwise convolutional operations, the depthwise separable convolution works on both spatial (i.e., height and width) and depth (i.e., number of channels) dimensions. From a structural point of view, the MBConv begins by expanding the number of channels by $1\text{×}1$ convolution, then employs depthwise convolution ($3\text{×}3$) and pointwise convolution ($1\text{×}1$), and finally the input and the output are added through the skip connection. In particular, the MBConv is utilized before the multi-axis attentions to improve the generalization and effective learning abilities, and the depthwise convolution enables the model to operate without explicit positional encoding. 
        
    On the other hand, the multi-axis attentions in MaxViT blocks basically consist of two types of attention mechanisms, such as block attention and grid attention to obtain local and global interactions, respectively. Let $\digamma_v \in \mathbb{R}^{w_{v}^{\prime} \times h_{v}^{\prime} \times c_{v}^{\prime}}$ is the intermediate feature map toward the attentions, and the shape of the blocks after partitioning the features into non-overlapping windows becomes $(w_{v}^{\prime}/\mathbf{w} \times h_{v}^{\prime}/\mathbf{w}, \mathbf{w} \times \mathbf{w}, \times c_{v}^{\prime})$, denoting $\digamma_{v\mathbf{w}}$ and each of ($\mathbf{w} \times \mathbf{w}$) size. Then, applying attention within each window independently, that is, the local spatial dimension ($\mathbf{w} \times \mathbf{w}$), one can obtain:
\begin{equation}
    \mathsf{Attention}(\mathbf{Q}_{v\mathbf{w}}, \mathbf{K}_{v\mathbf{w}}, \mathbf{V}_{v\mathbf{w}}) = \mathsf{Softmax}\left(\frac{\mathbf{Q}_{v\mathbf{w}}\mathbf{K}_{v\mathbf{w}}^{\mathsf{T}}}{\sqrt{d}} \right)\mathbf{V}_{v\mathbf{w}},
\end{equation}
    where, $\mathbf{Q}_{v\mathbf{w}} = \digamma_{v\mathbf{w}}\mathbf{W}_{v\mathbf{w}}^{Q}$, $\mathbf{K}_{v\mathbf{w}} = \digamma_{v\mathbf{w}}\mathbf{W}_{v\mathbf{w}}^{K}$, and $\mathbf{V}_{v\mathbf{w}} = \digamma_{v\mathbf{w}}\mathbf{W}_{v\mathbf{w}}^{V}$. And, we can get the multi-head attention within small local windows as: $\mathsf{Z}_{local} = \mathsf{MHA}(\mathbf{Q}_{v\mathbf{w}}, \mathbf{K}_{v\mathbf{w}}, \mathbf{V}_{v\mathbf{w}})$. 
    
    Following the block attention, the shape of the grids $\digamma_{v\mathbf{g}}$ is assumed as $(\mathbf{g} \times \mathbf{g}, w_{v}^{\prime}/\mathbf{g} \times h_{v}^{\prime}/\mathbf{g}, \times c_{v}^{\prime})$. The grid attention is then applied on a sparsely sampled uniform grid ($\mathbf{g} \times \mathbf{g}$), which can be calculated as:
\begin{equation}
    \mathsf{Attention}(\mathbf{Q}_{v\mathbf{g}}, \mathbf{K}_{v\mathbf{g}}, \mathbf{V}_{v\mathbf{g}}) = \mathsf{Softmax}\left(\frac{\mathbf{Q}_{v\mathbf{g}}\mathbf{K}_{v\mathbf{g}}^{\mathsf{T}}}{\sqrt{d}} \right)\mathbf{V}_{v\mathbf{g}},
\end{equation}
    where, $\mathbf{Q}_{v\mathbf{g}} = \digamma_{v\mathbf{g}}\mathbf{W}_{v\mathbf{g}}^{Q}$, $\mathbf{K}_{v\mathbf{g}} = \digamma_{v\mathbf{g}}\mathbf{W}_{v\mathbf{g}}^{K}$, and $\mathbf{V}_{v\mathbf{g}} = \digamma_{v\mathbf{g}}\mathbf{W}_{v\mathbf{g}}^{V}$. Next, the multi-head attention across non-overlapping grid stripes for long-range interactions across the entire image can be obtained by: $\mathsf{Z}_{global} = \mathsf{MHA}(\mathbf{Q}_{v\mathbf{g}}, \mathbf{K}_{v\mathbf{g}}, \mathbf{V}_{v\mathbf{g}})$. Note that the same fixed sizes ($\mathbf{w} = \mathbf{g} = 7$) are considered for both window and grid attentions, while the attention head size is set to be $32$, and a standard feed-forward network is applied after each attentions. 
    
    Based on this MaxViT block including MBConv, block attention, and grid attention, the overall MaxViT model builds a typical hierarchical structure by simply stacking four identical MaxViT blocks, however, each block has half resolution of the previous block along with a doubled channels. Finally, the outputs from the MaxViT blocks undergo a global average pooling layer, and in this way, the visual encoder helps to capture the detailed features from the visual inputs.
    
    \textit{3) Point Cloud Encoder:} For the point cloud encoder, we utilize the Point Transformer V3 (PTv3) model proposed in \cite{wu2024point} with the task-specific architectural modifications, motivated by the capability of extracting concise point cloud features in an efficient way in terms of faster speed and lower memory consumption. In particular, the PTv3 model is structured to work with four key components: (a) serialization, (b) grid pooling, (c) positional encoding, and (d) patch attention.
    
    The primary role of serialization is to take unordered set of points from point cloud $P = \{p_1, p_2, ..., p_{\mathsf{n}}\}$, where $P \in X^{\prime}_{p}$ and $p_i = \{(x_i, y_i, z_i)\}$, and serialize into a structured format $S = \{s_1, s_2, ..., s_{\mathsf{n}}\}$ so that it can fit with transformer style in subsequent operations. After applying the serialization, each point $s_i$ is mapped into a high-dimensional feature vector by an embedding function $\mathsf{Embed}(\cdot)$ as $\digamma_p = \mathsf{Embed}(s_i)$. The sequence of embeddings undergo further grid pooling and order shuffling processes.
    
    The PTv3 model is adopted for a patch attention mechanism. For that, the intermediate serialized point features $\digamma_p^{\prime} = \{s^{\prime}_1, s^{\prime}_2, ..., s^{\prime}_{\mathsf{n}}\}$ are first divided into non-overlapping patches (denoted as $\digamma_{p}^{\prime\prime}$), and the model then performs self-attention individually to each patch. For the $\mathsf{m}$-th patch, the attention can be derived as:
\begin{equation}
    \mathsf{Attention}(\mathbf{Q}_{p\mathsf{m}}, \mathbf{K}_{p\mathsf{m}}, \mathbf{V}_{p\mathsf{m}}) = \mathsf{Softmax}\left(\frac{\mathbf{Q}_{p\mathsf{m}}\mathbf{K}_{p\mathsf{m}}^{\mathsf{T}}}{\sqrt{d}} \right)\mathbf{V}_{p\mathsf{m}},
\end{equation}
    where, $\mathbf{Q}_{p\mathsf{m}} = \digamma_{p}^{\prime\prime}\mathbf{W}_{p\mathsf{m}}^{Q}$, $\mathbf{K}_{p\mathsf{m}} = \digamma_{p}^{\prime\prime}\mathbf{W}_{p\mathsf{m}}^{K}$, and $\mathbf{V}_{p\mathsf{m}} = \digamma_{p}^{\prime\prime}\mathbf{W}_{p\mathsf{m}}^{V}$. After that, we can get the returned output for $\mathsf{M}$ total number of patches as a sequence as $\digamma_{p}^{\mathsf{M}} = (f_1^{(1)}, ...,f_{\mathsf{n}}^{(1)}; f_1^{(2)}, ...,f_{\mathsf{n}}^{(2)}; ...; f_1^{(\mathsf{M})}, ...,f_{\mathsf{n}}^{(\mathsf{M})} )$. Additionally, an enhanced conditional positional encoding is added before the attention to work as a local attention while handling the voluminous point cloud samples. At the end, before feeding into feedforward network, a layer normalization is applied to the $\digamma_{p}^{\mathsf{M}}$ for faster convergence during training. 
    
\begin{figure*} [!t]
\centering
\begin{subfigure}[b]{0.24\textwidth}
    \centering
    \includegraphics[width=4.4cm]{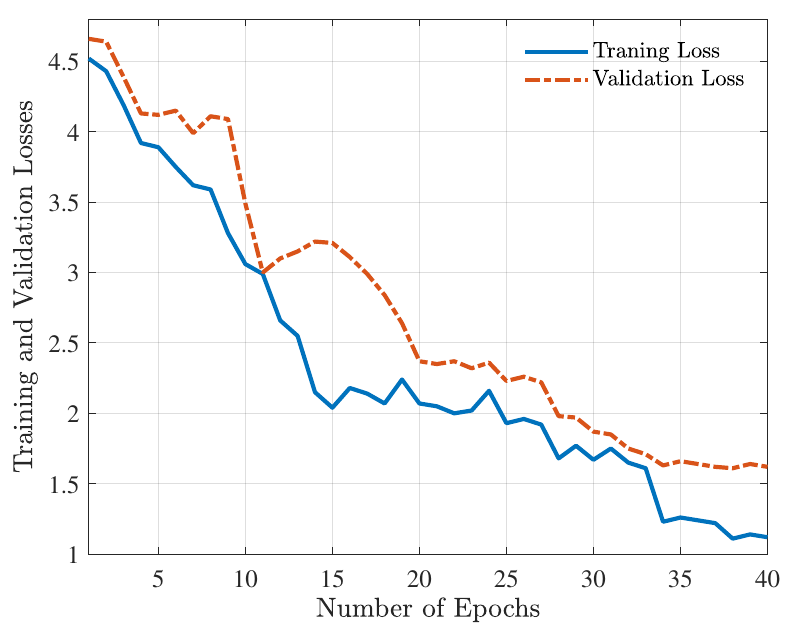}
\end{subfigure}
\begin{subfigure}[b]{0.24\textwidth}
	\centering
	\includegraphics[width=4.4cm]{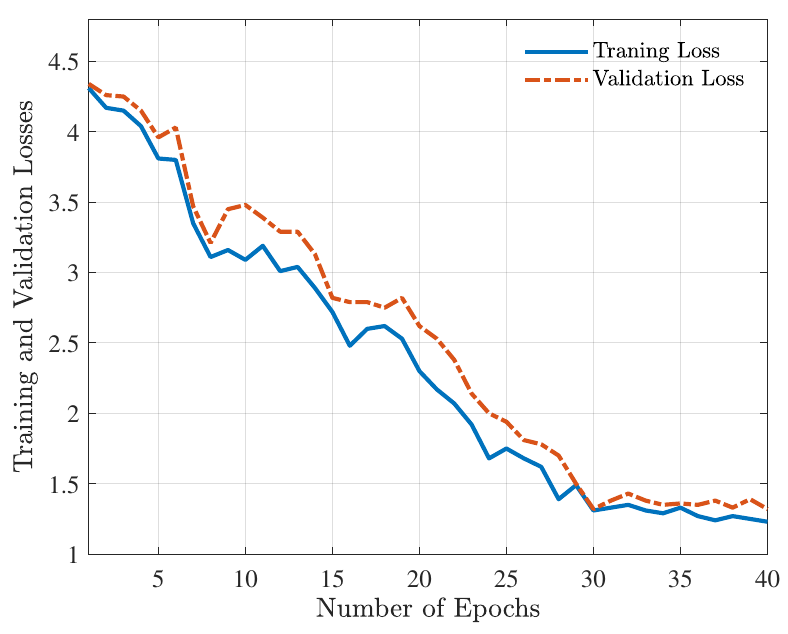}
\end{subfigure}
\begin{subfigure}[b]{0.24\textwidth}
    \centering
    \includegraphics[width=4.4cm]{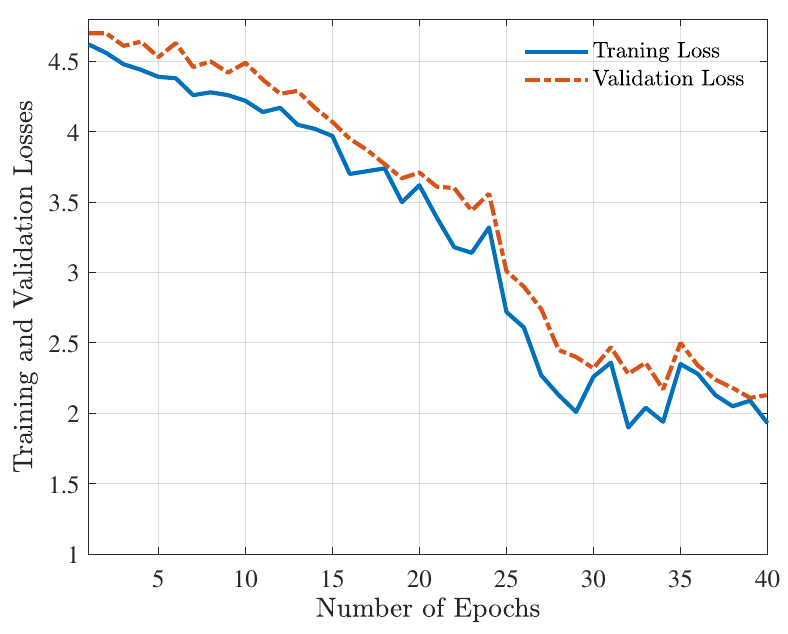}
\end{subfigure}
\begin{subfigure}[b]{0.24\textwidth}
    \centering
    \includegraphics[width=4.4cm]{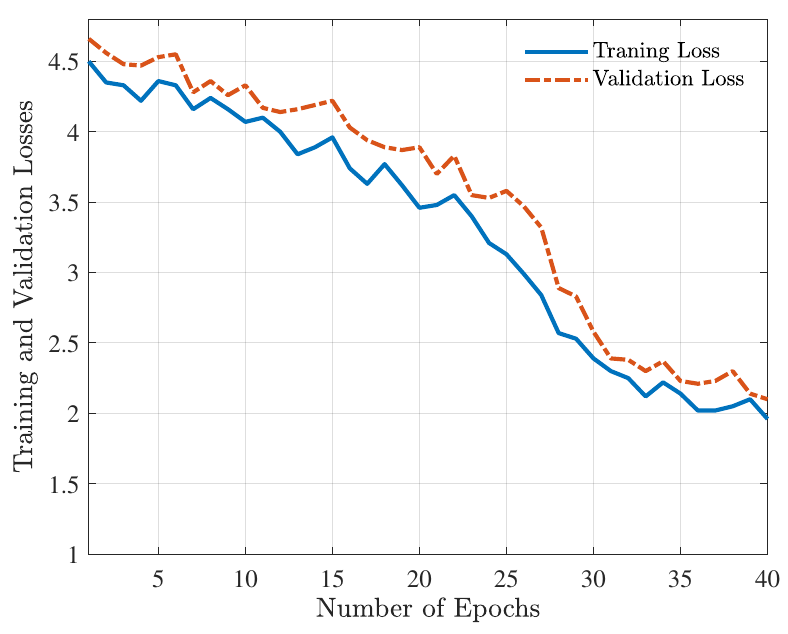}
\end{subfigure}
    \caption{The results on the basis of loss curves for V2I-Day, V2I-Night, V2V-Day, and V2V-Night scenarios (from left to right), assessing the the differences between predicted and ground truth beams during the training and validation processes over $40$ number of epochs.}
	\label{fig: losses}
\end{figure*}

\subsection{Fusion by Multi-Head Cross-Modal Attention}
    Following the aforementioned encoders within individual modalities, we include multi-head cross-modal attention between visual and point cloud modalities in our framework so that the dependencies and correlations from these modalities can be learned more effectively. Here, the visual modality can act as the query while another point cloud modality can function as the key-value pair and vice versa. For that, let the respective outputs from each encoders be denoted as $\mathsf{Z}_{v}$ and $\mathsf{Z}_{p}$. Aiming to perform cross-modal multi-head attention between visual and point cloud features, we first calculate the cross-modal attention from $\mathsf{Z}_{p}$ to $\mathsf{Z}_{v}$ as:
\begin{equation}
    \mathsf{CrossAtt}(\mathbf{Q}_{\mathsf{Z}_{v}}, \mathbf{K}_{\mathsf{Z}_{p}}, \mathbf{V}_{\mathsf{Z}_{p}}) = \mathsf{Softmax}\left(\frac{\mathbf{Q}_{\mathsf{Z}_{v}}\mathbf{K}_{\mathsf{Z}_{p}}^{\mathsf{T}}}{\sqrt{d}} \right)\mathbf{V}_{\mathsf{Z}_{p}}.
\end{equation}
    Here, the corresponding query, key, and value are calculated by $\mathbf{Q}_{\mathsf{Z}_{v}} = \mathsf{Z}_{v}^{\prime}\mathbf{W}_{\mathsf{Z}_{v}}^{Q}$, $\mathbf{K}_{\mathsf{Z}_{p}} = \mathsf{Z}_{p}^{\prime}\mathbf{W}_{\mathsf{Z}_{p}}^{K}$, and $\mathbf{V}_{\mathsf{Z}_{p}} = \mathsf{Z}_{p}^{\prime}\mathbf{W}_{\mathsf{Z}_{p}}^{V}$, where $\mathsf{Z}_{v}^{\prime} = \mathsf{Z}_{v}\mathbf{W}_{v} + \mathbf{b}_{v}$ and $\mathsf{Z}_{p}^{\prime} = \mathsf{Z}_{p}\mathbf{W}_{p} + \mathbf{b}_{p}$ are the embedded features, each having same $d_z$ dimensions. In a similar manner, we also calculate the cross-modal attention from $\mathsf{Z}_{v}$ to $\mathsf{Z}_{p}$ as $\mathsf{Attention}(\mathbf{Q}_{\mathsf{Z}_{p}}, \mathbf{K}_{\mathsf{Z}_{v}}, \mathbf{V}_{\mathsf{Z}_{v}})$ considering the interactions between visual and point cloud features are mutual. Then, we can get the heads as follows:
\begin{equation}
    head^{(h)}_{p \rightarrow v} = \mathsf{CrossAtt}(\mathbf{Q}^{(h)}_{\mathsf{Z}_{v}}, \mathbf{K}^{(h)}_{\mathsf{Z}_{p}}, \mathbf{V}^{(h)}_{\mathsf{Z}_{p}})
\end{equation}
\begin{equation}
    head^{(h)}_{v \rightarrow p} = \mathsf{CrossAtt}(\mathbf{Q}^{(h)}_{\mathsf{Z}_{p}}, \mathbf{K}^{(h)}_{\mathsf{Z}_{v}}, \mathbf{V}^{(h)}_{\mathsf{Z}_{v}}),
\end{equation}
    where, we set number of heads $h = 8$, and following these two heads, we can construct the individual multi-head cross modal attentions as:
\begin{equation}
    \digamma_{p \rightarrow v} = \mathsf{CrossMHA}(\mathsf{Z}_{v}^{\prime}\mathbf{W}_{\mathsf{Z}_{v}}^{Q{(h)}}, \mathsf{Z}_{p}^{\prime}\mathbf{W}_{\mathsf{Z}_{p}}^{K{(h)}}, \mathsf{Z}_{p}^{\prime}\mathbf{W}_{\mathsf{Z}_{p}}^{V{(h)}})
\end{equation}
\begin{equation}
    \digamma_{v \rightarrow p} = \mathsf{CrossMHA}(\mathsf{Z}_{p}^{\prime}\mathbf{W}_{\mathsf{Z}_{p}}^{Q{(h)}}, \mathsf{Z}_{v}^{\prime}\mathbf{W}_{\mathsf{Z}_{v}}^{K{(h)}}, \mathsf{Z}_{v}^{\prime}\mathbf{W}_{\mathsf{Z}_{v}}^{V{(h)}}).
\end{equation}
    
    After the individual multi-head cross modal attentions processing, the resultant attentions $\digamma_{p \rightarrow v}$ and $\digamma_{v \rightarrow p}$ are concatenated to obtain final desired bidirectional relationships, followed subsequently by layer normalization to facilitate feature stability as:
\begin{equation}
    \mathsf{Z}_{v,p} = \mathsf{LayerNorm}\left(\mathsf{Concet}(\digamma_{p \rightarrow v}, \digamma_{v \rightarrow p})\right).
\end{equation}
    
\subsection{Top-$k$ Beam Selection}
    Given $\mathsf{Z}_g$ from the position encoder and $\mathsf{Z}_{v,p}$ from the fusion of visual and point clouds by multi-head cross-modal attention, the beam selection module produces the final prediction results. For that, the feature fusion along their feature dimensions through a concatenation function is first carried out to obtain a unified representation. The resulting concatenated output is then forwarded to a multi-layer perceptron network, which is basically made of Rectified Linear Unit (ReLU) activation function and layer normalization procedures at each layer. We can describe the overall processes as:
\begin{equation}
    \begin{array}{l}
    \mathsf{Z}_{fused} = \mathsf{Concet}(\mathsf{Z}_g, \mathsf{Z}_{v,p}) \\
    \mathsf{MLP}_1 = \mathsf{ReLU}(\mathsf{LayerNorm}(\mathsf{Z}_{fused}\mathbf{W}_1 + \mathsf{b}_1)) \\
    \mathsf{MLP}_2 = \mathsf{ReLU}(\mathsf{LayerNorm}(\mathsf{MLP}_1\mathbf{W}_2 + \mathsf{b}_2)) \\
    \hat{\mathsf{b}} = \mathsf{Softmax}(\mathsf{MLP}_2\mathbf{W}_3 + \mathsf{b}_3),
    \end{array}
\end{equation}
    where, $\mathbf{W}$ and $\mathsf{b}$ are the corresponding weights and biases, which will be learned during training. At the final layer, the $\mathsf{Softmax}(\cdot)$ function is applied to make a normalized output into a probabilistic prediction, which can be essentially interpreted to determine the $k$ best beams.
    
\section{Experiments and Performance Assessment}
    In this section, we present the considered dataset and experimental setup in details. This is followed by a discussion of the performance comparison through a series of extensive experiments to show the effectiveness of the proposed work over the state-of-the art baselines.

\subsection{Dataset Overview}
    For training and performance evaluation, we consider the DeepSense 6G dataset \cite{alkhateeb2023deepsense, morais2025deepsense}, which is a collection of coexisted multi-modal sensor observations captured from real-world wireless environments. Consistent with the considered system model, the scenarios 31, 33, 36, and 37 from this dataset are particularly adopted for experimentation. Here, the sensing observations of scenarios 31 (day) and 33 (night) are collected under vehicle-to-infrastructure (V2I) communications, whereas the observations of scenarios 36 (day) and 37 (night) are collected under vehicle-to-vehicle (V2V) communications, with $7,012$, $3,981$, $24,800$, and $31,000$ number of samples, respectively. For each V2I and V2V scenarios, two separate testbed setups with transmitter and receiver units operating at 60 GHz bands are implemented to collect data at different locations, such as Tempe, Phoenix, and Scottsdale, in Arizona.
    
    In particular, the mmWave phased arrays at receiver units employ $16$-element ($N_{rx} = 16$) phased array antennas with an over-sampled $64$ beamsteering codebooks ($|\mathbfcal{C}_{rx}| = 64$), while the transmitter units have one antenna element ($N_{tx} = 1$) to make the omnidirectional transmission towards the receivers. At each discrete time intervals, the receiver units captures the following samples (i) the visual and point clouds by scanning the surrounding environment, (ii) the GPS coordinates from the transmitters, (iii) the received powers at each beam codebooks representing as the sequence indices as $\mathcal{B}_{rx} \in \{1, 2, 3, ...., 64\}$ through a full beam sweeping approximately from $-45$-degree to $+45$-degree in azimuth, and (iv) the ground truth beams, that is, the optimal beam indices computed from the received powers. Most importantly, the captured samples are synchronized across all collected samples and recorded at $0.1$ second of time interval (i.e., $10$ samples/second).
    
\subsection{Experimental Setup and Network Training}
    In this work, a series of experiments were conducted by implementing the designed multi-modal model with the considered dataset on an Intel Core i7-10875H CPU and NVIDIA GeForce RTX 2080 Super GPU based computing platform. Specifically, we use Pytorch 1.13.1 framework in Python 3.7 and CUDA toolkit 11.7 to develop the model on this platform. During training, the adaptive moment estimation (Adam) \cite{KingBa15} optimizer was utilized with a learning rate of $0.001$ and weight decay of $1 \times 10^{-4}$ for faster convergence while minimizing the loss function. 
    
    For the purpose of experimentation, we randomly partition the data resources of each scenario into training, validation, and testing sets using a ratio of 6:2:2. Following these partitioning, the model is trained over $40$ epochs with a batch size of $4$ for each to get the decent accuracy. Fig. \ref{fig: losses} illustrates the results of the training and validation losses obtained at each epoch for the considered scenarios, and from the results, one can observe that both losses minimize and stabilize at a certain point, leading to a perfect fitting. The codes and other related instructions are publicly available at GitHub\footnote{\url{https://github.com/mbaqer/V2X-mmWave-Transformers/}}. 
    
\begin{figure*} [!t]
\centering
\begin{subfigure}[b]{0.24\textwidth}
    \centering
    \includegraphics[width=4.4cm]{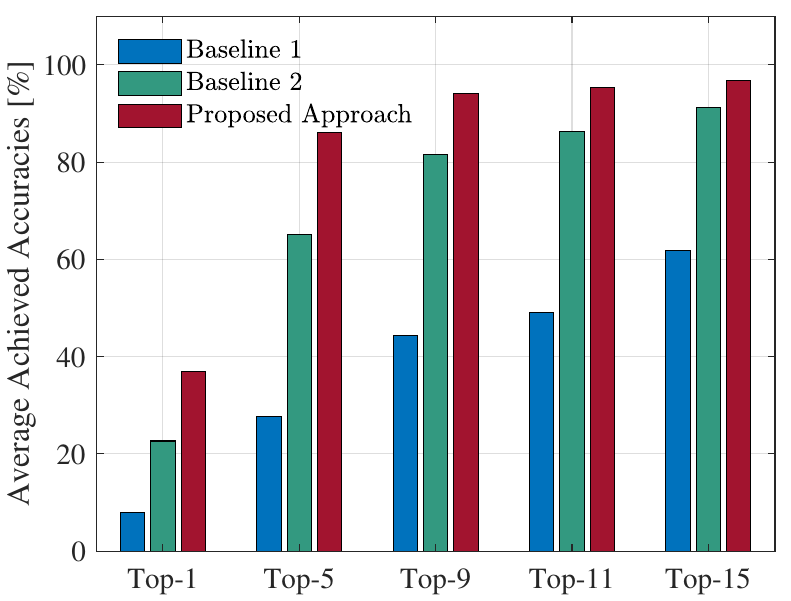}
	\caption{\centering \scriptsize For V2I-Day Scenario}
\end{subfigure}
\begin{subfigure}[b]{0.24\textwidth}
	\centering
	\includegraphics[width=4.4cm]{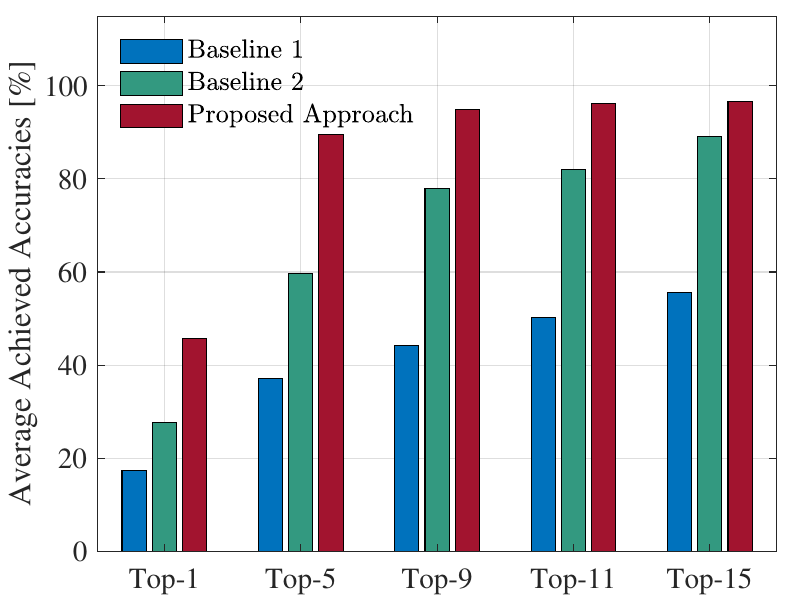}
	\caption{\centering \scriptsize For V2I-Night Scenario}
\end{subfigure}
\begin{subfigure}[b]{0.24\textwidth}
    \centering
    \includegraphics[width=4.4cm]{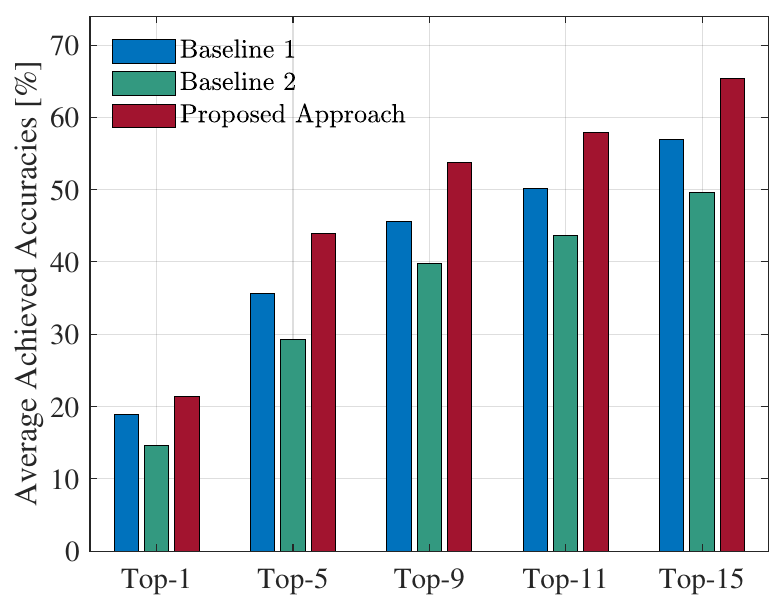}
    \caption{\centering \scriptsize For V2V-Day Scenario}
\end{subfigure}
\begin{subfigure}[b]{0.24\textwidth}
    \centering
    \includegraphics[width=4.4cm]{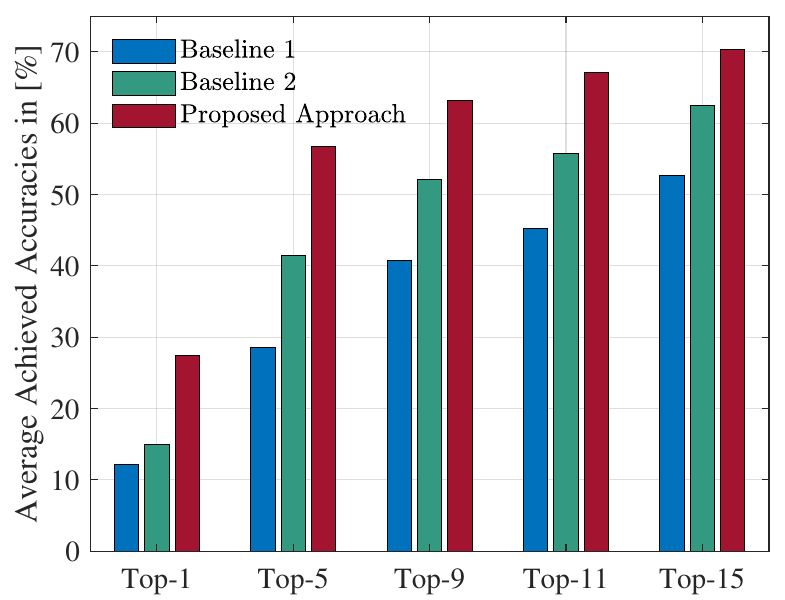}
    \caption{\centering \scriptsize For V2V-Night Scenario}
\end{subfigure}

\vspace{3mm}

\begin{subfigure}[b]{0.24\textwidth}
    \centering
    \includegraphics[width=4.4cm]{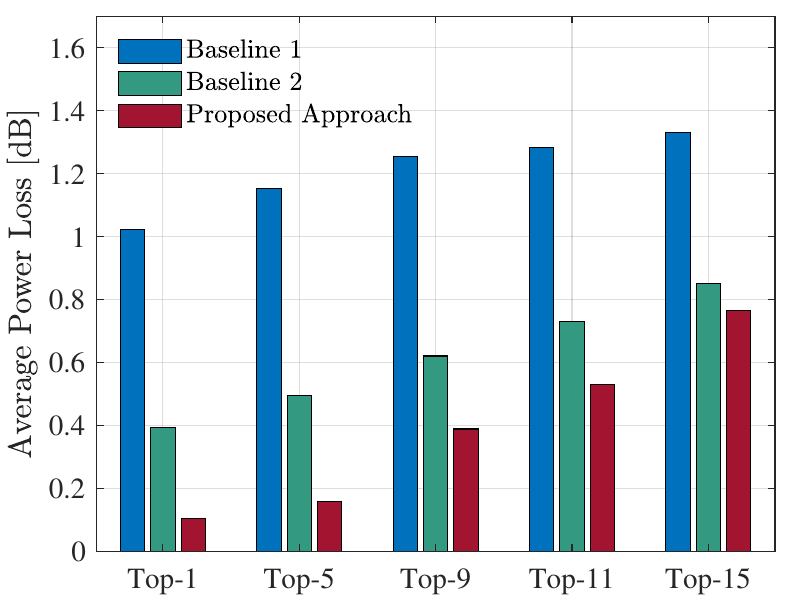}
	\caption{\centering \scriptsize For V2I-Day Scenario}
\end{subfigure}
\begin{subfigure}[b]{0.24\textwidth}
	\centering
	\includegraphics[width=4.4cm]{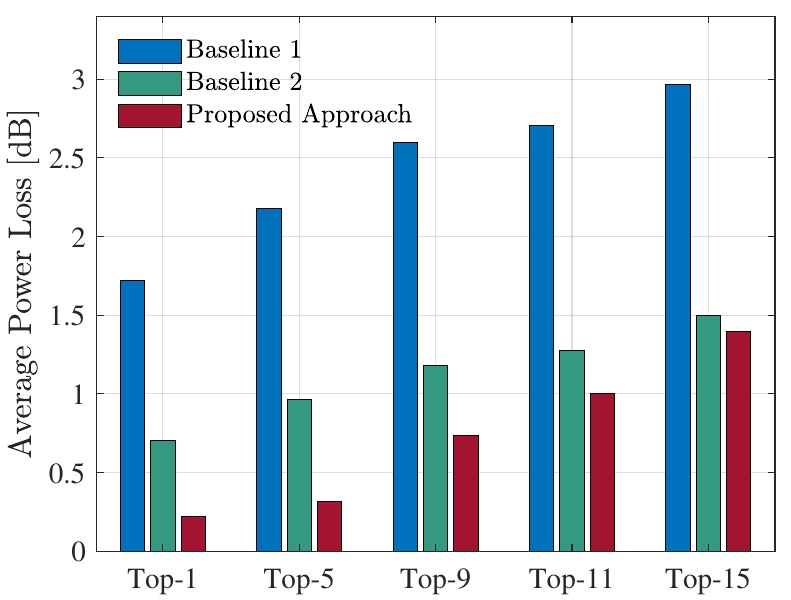}
	\caption{\centering \scriptsize For V2I-Night Scenario}
\end{subfigure}
\begin{subfigure}[b]{0.24\textwidth}
    \centering
    \includegraphics[width=4.4cm]{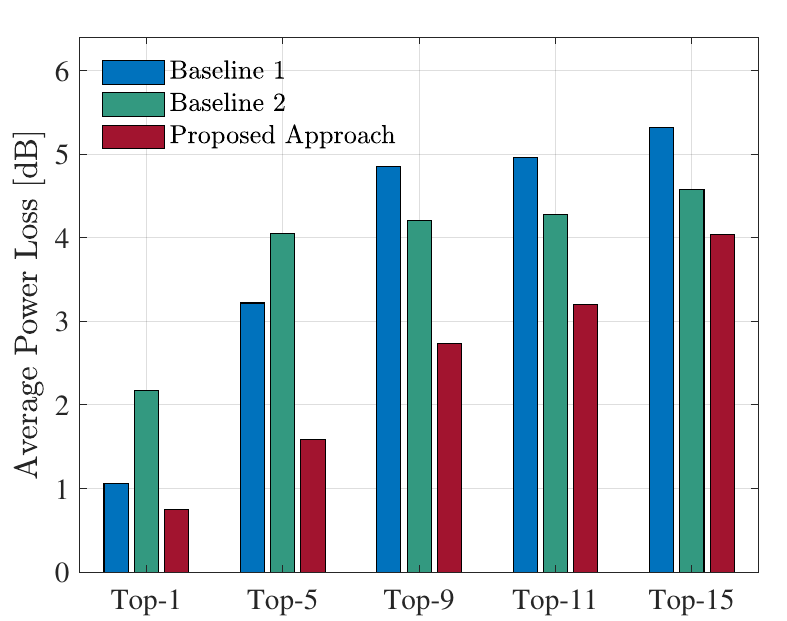}
    \caption{\centering \scriptsize For V2V-Day Scenario}
\end{subfigure}
\begin{subfigure}[b]{0.24\textwidth}
    \centering
    \includegraphics[width=4.4cm]{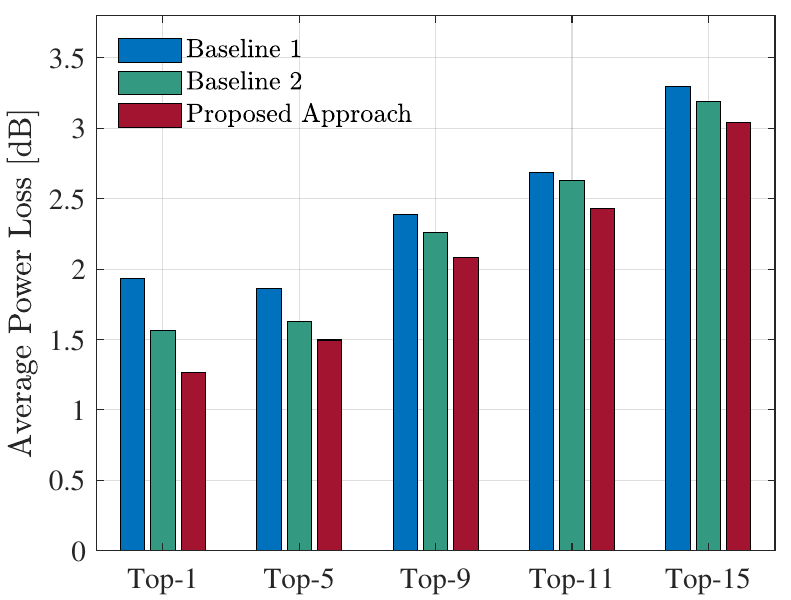}
    \caption{\centering \scriptsize For V2V-Night Scenario}
\end{subfigure}
    \caption{The performance comparisons on average achieved accuracies in percentages and average power mmWave loss in decibels tested on all considered V2I and V2V scenarios.}
	\label{fig: Performances}
\end{figure*}

\subsection{Performance Assessment}
    Upon completing the training processes, the following two key metrics are utilized to assess the performance of the proposed model.
    
    \textit{1) Top-$k$ Accuracy:} is defined as the number of correct top-$k$ most likely outcomes with respect to true beams by the model. This metric essentially presents the model's ability on making accurate predictions. Defining the top-$k$ accuracy as $Acc_{top-k}$, it can be given by:
\begin{equation}
    Acc_{top-k} = \frac{1}{N_{test}}\sum_{n=1}^{N_{test}} \left[\mathbbm{1}(\hat{\mathsf{b}}_n = Y^{gt}_n)\right],
\end{equation}
    where, $\mathbbm{1}(\cdot)$, $N_{test}$, $\hat{\mathsf{b}}_n$ and $Y^{gt}_n$ are the indicator function, total number of test samples, predicted beams at $n$-th test sample, ground-truth beams at $n$-th test sample, respectively.
    
    \textit{2) Average Power Loss:} is calculated from the average ratio between the downlink received powers with top-$k$ predicted beams and the maximum received powers possibly achieved at the link while considering the noises in realistic scenarios. In decibels (dB) scale, it can be expressed as:
\begin{equation}
    APL[\text{dB}] = - 10 \log_{10} \left( \frac{1}{N_{test}}\sum_{n=1}^{N_{test}}\frac{\hat{\mathsf{p}}_n-\mathsf{p}_o}{\mathsf{p}_n^{gt}-\mathsf{p}_o}\right),
\end{equation}
    where, $\hat{\mathsf{p}_t}$, $\mathsf{p}_t^{gt}$, $\mathsf{p}_o$ are the corresponding powers from predicted beams at the $n$-th test sample, the corresponding powers ground-truth beams at the $n$-th test sample, and noise powers, respectively.
    
    With these performance metrics, we perform a comparative study by evaluating the performance assessment of the proposed mode with two established approaches as baselines. In the first baseline, we adopt a uni-modal data based approach. Similar to \cite{morais2023position}, this approach utilizes only a single GPS sensing coordinates modality, hence no fusion, to predict the most-likely beams. However, we have particularly used the position data feature extraction module part from the proposed multimodal model. This baseline approach is considered to confirm how adding more modalities by fusion can help to enhance the prediction accuracy. Further, for the second baseline, the fusion model presented in \cite{fabiani2024multi} is adopted. This fusion model basically lets each extracted features from modalities concatenate them straightforwardly before passing through a multi-layer perceptron with multiple layers.

    Fig. \ref{fig: Performances} illustrates average achieved accuracies (in percentage) as well as average power loss (in dB) for the proposed approach when compared with both of the baselines across all considered V2I-day, V2I-night, V2V-day, and V2V-night scenarios. Here, each of the x-axis presents the top beam candidates as top-$k$, where $k \in \{1, 5, 9, 11, 15\}$. Specifically, the average achieved accuracies results for the model as in Fig. \ref{fig: Performances}(a)-(d) show that the baseline 1 on single modality (that is, without fusing with other modalities) performs the worst among others in most of the scenarios, essentiality leading to many wrong beam directions. Notably, after performing the multi-modal fusion, the performance for baseline 2 gets better than that for baseline 1 as expected. In contrast, the proposed fusion-based approach with cross-modality multi-head attention explicitly outperforms over both baseline 1 and baseline 2 across all considered scenarios. For example, the V2I-day scenario achieves the maximum beam prediction accuracy of $96.72$\% with the proposed approach while predicting the top $15$-beams, whereas the baseline 1 and baseline 2 attain accuracy of $61.82$\% and $91.31$\%, respectively. However, the lower accuracy in V2V is mainly because vehicle-to-vehicle links are more dynamic, with frequent changes in positions, blockages, and channel conditions, making beam prediction more challenging compared to the relatively stable vehicle-to-infrastructure links.

    Consequently, we can further observe in Fig. \ref{fig: Performances}(e)-(h) that the proposed solution on all considered four scenarios experience a significant reduction in terms of the average power losses with respect to the baseline 1 and baseline 2. For instance, the proposed approach incurs only $0.77$~dB average power loss in V2I-day scenario with top $15$ beams, compared to $1.33$~dB and $0.85$~dB respectively for baseline 1 and baseline 2. Most importantly, this metric enables us to accurately represent the actual impact of the harnessed power from the predicted beams particularly by quantifying how much powers might be lost when the predicted beams are used. In other words, better energy efficiency inherently comes with lower power loss. Hence, such performance results indicates that the proposed approach offers also promising capability of offering more energy efficient beamforming solution.
        
    In summary, taking advantage of multi-head cross-modality attention mechanism in fusion model enables learning the features from the modalities as well as mapping between the modalities and beams more effectively. Consequently, the proposed approach leads to improved prediction accuracy and lowered power loss. On the other hand, the dataset used in this experiment is based on publicly available measured data collected in urban environments during both day and night times. However, for real world applicability, it is also important to access how this work can be adaptable to suburban or rural scenarios. Furthermore, the sensing information might be impacted due to different weather conditions, which makes them an important factor to be considered. Thus, we suggest updating and fine-tuning the trained model occasionally while ensuring the minimal performance loss might be necessary to improve the generalization ability for real-world deployment.

    Besides, the dataset we have used primarily provides mappings between beam codebooks and measured received powers under line-of-sight (LoS) conditions for both V2I and V2V scenarios. Other multi-modal sensing datasets, such as FLASH \cite{flash}, do not consider V2V settings, limiting their applicability in this work. Hence, it should be noted that since our proposed framework already utilizes multi-modal data, it could be further fine-tuned and revised to predict non-line-of-sight (NLoS) or unseen scenarios and adjust beamforming strategies accordingly. Nevertheless, a potential direction for adapting the proposed multi-model to NLoS or unseen scenarios is to employ incremental learning strategy, where the model can be updated using small amounts of newly collected data without requiring full re-training. In particular, this can be achieved through lightweight parameter updates, such as using freezing and selective fine-tuning of transformer layers. In practice, once vehicles encounter a new scenario, such as different road geometry, weather, or propagation conditions, the system is expected to incorporate these samples into a continual learning pipeline to gradually improve its generalization while preserving performance on previously seen environments. Eventually, this enables the model to remain adaptive and robust in real-world deployments where operating conditions evolve over time.
    
\section{Integration with 5G-NR}
    In this section, we present a comprehensive overview of the existing beam management procedure according to 5G-NR (New Radio) standard, and how the proposed approach can be integrated with it. Finally, the applicability of the proposed solution in practical environment is explored considering the analysis of communication and computational overheads.

\subsection{Beam Management in 5G-NR}
    Beam management is essential for mmWave communications to establish and maintain highly directional communication links with satisfactory signal strength and reduced interference. Typically, the beam management procedure involves two key steps. The first step, usually referred to as initial access, where the transmitter and receiver nodes perform initial beam selection by discovering the beam directions with sufficient signal strength. The signal strength can be estimated in terms of received power or signal-to-noise-ratio (SNR). Following the initial beam establishment, both nodes perform beam tracking to avoid sudden signal strength drops by maintaining the selected beams perfectly aligned as time progresses.
    
\begin{figure*} [!t]
	\includegraphics[width=.90\linewidth]{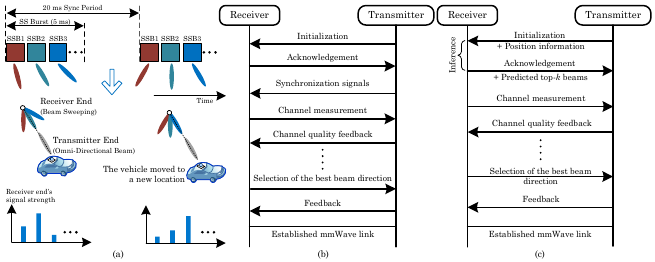}
    \centering
    \caption{(a) The 5G-NR standard defined frame structure of synchronizing signal (SS) burst (SSB1, SSB2, etc. within an SS burst in different colors are the SS blocks associated with each beam) and (b)\&(c) the procedures of standardized defined exhaustive beamforming and after integrating the proposed multimodal model into it, represented as timing diagrams.}
    \label{fig: Timing_5G_Proposed}
\end{figure*}
    
    Toward this goal, the 5G-NR standard defines a codebook-based approach for beam management for both initial access and beam tracking \cite{tan2024beam,xue2024ai}. In such beam management procedure as depicted in Fig. \ref{fig: Timing_5G_Proposed}(a), the transmitter and receiver nodes coordinate with each other by exchanging pilot signals and channel feedback. For that, the standard adopts a sequence of synchronization signal blocks (SS blocks), where each SS block is assigned for a specific beam direction. To measure the beam quality, a set of SS blocks are first sent by the transmitter sequentially to each of its beam directions. At the same time, the receiver unit listens under each directions and this process is repeated with reverse role if required. In particular, the set of SS blocks together are referred to as a single burst with a maximum $5$ ms time length, and a single burst is time-multiplexed of such set of SS blocks. Besides, each single burst repeats every $20$ ms (default) periodically to cover the full exploration among all possible beam directions. In this way, the transmitter and receiver exhaustively explore by sweeping all possible codebook of directions with a brute-force-based way to determine the best beams providing the highest beamforming gains.
    
\subsection{Details of Integration}
    Before discussing the details of integration, we first present the working aspects for better understanding as a timing diagram for the 5G-NR standard defined exhaustive beam searching approach as shown in Fig. \ref{fig: Timing_5G_Proposed}(b). In general, the beam search process is initialized by the transmitter by sending a beam training request to the receiver. Upon receiving the acknowledgment, the transmitter begins the major procedure, that is, beam sweeping. During this procedure, the receiver/transmitter nodes exchange the synchronization sequences and quality of the received signals within the spatial area. However, the beam sweeping resumes after the burst period ($20$ ms) again in case the full exploration is not completed within first burst. In the end, the transmitter/receiver make an informed decision from the quality of the received signals on the best beam(s) to be used for directional communication for the initial access and subsequent beam trackings.
    
    Fig. \ref{fig: Timing_5G_Proposed}(c) illustrates the procedures on how the proposed beamforming approach will work after integrating with the 5G-NR standard defined approach. In particular, the transmitter vehicle who wish to establish a mmWave communication link first sends the beam training request along with its position information to the receiver road-side unit (V2I) or the other connected vehicle (V2V). After that, the receiver utilizes the position information along with other vision modalities to infer the top-$k$ candidate beams with the help of trained model. Finally, the beam training follows the rest of standardized procedures, where the beam sweeping for searching the accurate beam direction will be performed only over the predicted top-$k$ beams. However, it is worth mentioning that the top-$1$ predicted beams may lead to incorrect beam selection that yields poor link quality, even though the top-$1$ is expected to be the optimal beam. This is primarily because the prediction accuracy of top-$1$ is relatively lower than that of the top-$15$. Thus, the simple intuition is utilizing the predicted 15-beams would perform better, thereby reducing the possibility of beam misalignment.
    
\subsection{Latency and Beam Searching Overheads}
    Section V presents how the proposed work outperforms other state-of-the art baselines in terms of top-$k$ accuracy and average power loss, however, we now calculate the end-to-end latency and beam searching overheads as per 5G-NR standard. For calculating the end-to-end latency, we adopt the analysis from \cite{salehi2024omni} and consider each synchronization signal (SS) burst have up to $32$ blocks, which allows for sweeping $32$ maximum beams within a burst. Defining $t_{bs}$, $T_{ssb}$, and $K$ correspond respectively to the SS burst duration, its periodicity, and the total number of beams in codebook, one can express the time associated with the beam sweeping process as:
\begin{equation}
    T_{sp}(K) = T_{ssb} \times \left\lfloor\dfrac{K-1}{32}\right\rfloor + t_{bs},
    \label{eq:5G-NR_Eqn_1}
\end{equation}
    where, $t_{bs}$ = $5$ ms and $T_{ssb}$ = $20$ ms (default) as per the 5G-NR standard specification. Then, with the $5$~ms single burst time, we can get the time for exploring one single beam as $t_{ssb}$ = $t_{bs}/32$ = $0.156$ ms. However, given the aim of the proposed approach on seeking to reduce the number of beams to be swept from $K$ to most likely top-$k$ candidates, the beam sweeping time for the top-$k$ predicted beams can be given by:
\begin{equation}
    T_{sp}^{mm}(k) = T_{ssb} \times \left\lfloor\dfrac{k-1}{32}\right\rfloor + t_{ssb}(1+(k-1)\;\mathrm{mod}\;32).
    \label{eq:5G-NR_Eqn_2}
\end{equation}
    
    As a result, the total time required to complete the beam selection process can be estimated to be $\mathcal{T}(k) = T_{prc} + T_{inf} + T_{fb} + T_{sp}^{mm}(k)$. Here, the first three terms represent the required times for multi-modal data preprocessing, inference, and feedback, respectively. After measuring the inference time for the experiment done on NVIDIA GeForce RTX 2080 GPU with the testing set, the average per sample inference execution time falls in-between $0.5$ ms and $1.0$ ms for the average per sample inference execution time, we thus set $T_{inf}$ = $1.0$~ms. Besides, the data preprocessing and feedback times are typically very small, as a result, together we have $T_{prc} + T_{fb}$ $\le$ $0.1$ ms.
    
\begin{figure} [!ht]
	\includegraphics[width=.75\linewidth]{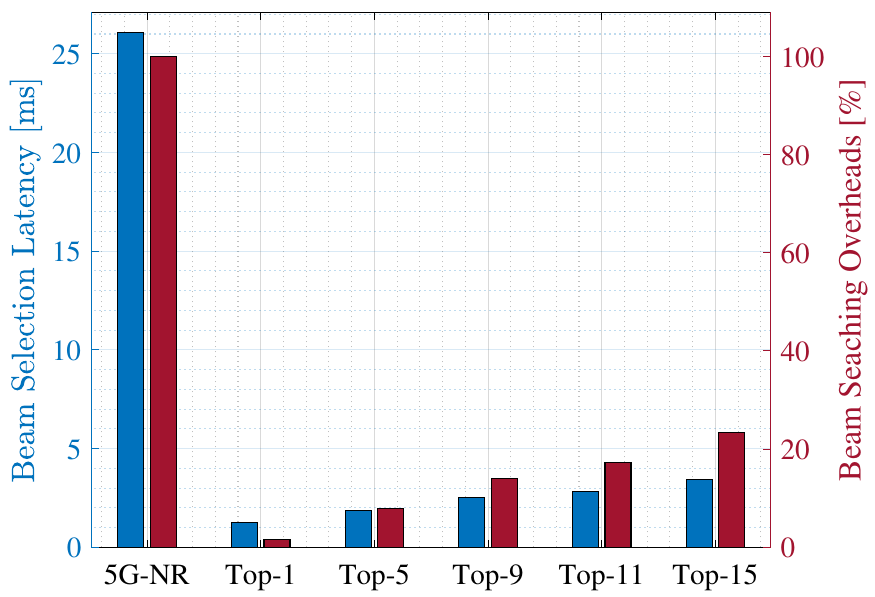}
    \centering
    \caption{The performance results considering beam selection and searching overheads for existing 5G-NR based approach with the proposed solution.}
    \label{fig: Latency_Overheads}
\end{figure}

    Fig. \ref{fig: Latency_Overheads} illustrates the performance comparison of the proposed approach with 5G-NR defined exhaustive search based beam selection approach with focus on the end-to-end latency and searching space overheads. In particular, the Eqs. (\ref{eq:5G-NR_Eqn_1}) and (\ref{eq:5G-NR_Eqn_2}) are used to calculate the end-to-end latency. As presented, roughly $3.44$ ms time will be required if we choose to limited the exploring among the predicted 15-beams, which considerably shortens from the increased latency $26.1$ ms by the exhaustive search. In contrast, in line with this work, the beam searching overheads solely rely on the number of beams to be explored. Accordingly, the proposed multi-modal model provides predicted beam candidates leading to a reduction of $\text{top-}k/K$ beam search space overheads. For instance, in case of the searching the best beam from the predicted top-$15$ beams, the beam searching overheads can be reduced to $23.43\%$ of the total $64$ beams.

    Overall, we recall that the well-known limitations of the exhaustive search defined in 5G-NR are the sweeping overheads that scales along with the number of antennas as well as the latency introduced by synchronization signal transmission, beam measurements, and uplink/downlink feedbacks. In particular, we can observe from the aforementioned results that both beam selection latency and large beam sweeping overheads have been reduced substantially when switching from the conventional exhaustive approach to that of the current work. For example, utilizing multi-modal sensory information for beam selection has helped to shorten the latency and searching space overheads by $86.81$\% and $76.56$\% for $15$-beams, at the same time, the proposed model has up to $96.72$\% accuracy on predicting correctly with $0.77$~dB on average power loss as observed in earlier section. For that reason, in spite of taking longer time for the model training, measuring the average inference execution time on NVIDIA GeForce RTX 2080 GPU, ranges between $0.5$ ms and $1.0$ ms (as mentioned earlier). Such results reaffirms the applicability of the proposed work in full-filling the requirement while deploying in practical mmWave enabled vehicular networks.
    
\section{Conclusion}
    In this paper, we have presented a 5G-NR standard compatible beam selection solution for $60$ GHz mmWave enabled connected vehicles. Based on the out-of-band contextual sensing information observed from the surroundings along with position information, we have designed a transformer based multi-modal fusion framework to predict top-$k$ beams, that is a subset of beams. Subsequently, we have evaluated the proposed framework with practically collected wireless datasets in diverse scenarios. In particular, our findings have highlighted the competitive performances of the proposed framework in terms of prediction accuracies, average power losses, and beam training overheads, marking the ability to meet high throughput and low latency demands of connected vehicles. Perhaps, exploring how to incorporate with emerging multimodal foundation models could be considered as a future work.
    
\ifCLASSOPTIONcaptionsoff
  \newpage
\fi
    
\bibliographystyle{IEEEtran}
\bibliography{References.bib}

\begin{IEEEbiography}[{\includegraphics[width=1in,height=1.25in,clip,keepaspectratio]{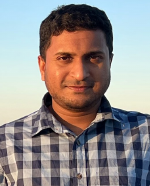}}]
    {Muhammad Baqer Mollah} received his PhD from the Dept. of Electrical and Computer Engineering at the University of Massachusetts Dartmouth. He is currently working as a post-doctoral fellow at University of Houston. His research interests include advanced sensing, communication, security, and edge intelligence techniques for cyber-physical systems, such as transportation and mobility, smart grids, industrial systems, and smart cities, among others. His works have been published in major IEEE Transactions, Magazines, and Conferences. He has a M.Sc. in Computer Science and B.Sc. in Electrical and Electronic Engineering from Jahangirnagar University, Dhaka and International Islamic University Chittagong, Bangladesh, respectively.
\end{IEEEbiography}

\begin{IEEEbiography}[{\includegraphics[width=1in,height=1.25in,clip,keepaspectratio]{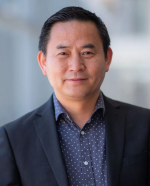}}]
    {Honggang Wang} is the department chair and Professor of the Dept. of Graduate Computer Science and Engineering at Katz School of Science and Health, Yeshiva University. He was a Professor in the Department of Electrical and Computer Engineering, University of Massachusetts Dartmouth. He was the “Scholar of The Year” awardee in 2016, the highest research recognition at University of Massachusetts Dartmouth. He is currently the Editor-in-Chief of IEEE Transactions on Multimedia and has served as the Editor-in-Chief of IEEE Internet of Things Journal (2020 - 2022). His recently focused research topics involve AI and Internet of Things, mmWave communications, connected vehicles, smart health, and cyber security. He has a PhD in Computer Engineering from University of Nebraska - Lincoln.
\end{IEEEbiography}

\begin{IEEEbiography}[{\includegraphics[width=1in,height=1.25in,clip,keepaspectratio]{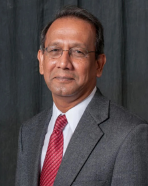}}]
    {Mohammad Ataul Karim} is a Professor in the Dept. of Electrical and Computer Engineering at the University of Massachusetts Dartmouth. His areas of research encompass optical computing, pattern/target recognition, computer vision, and Internet of Things. He is the former Provost, Executive Vice Chancellor for Academic Affairs, and Chief Operating Officer at University of Massachusetts Dartmouth. He received his PhD in electrical engineering degree from the University of Alabama. Karim is a fellow also of Optica, Society of Photo-Instrumentation Engineers, Institute of Physics, Institution of Engineering and Technology, Bangladesh Academy of Sciences, and Asia-Pacific Artificial Intelligence Association.
\end{IEEEbiography}

\begin{IEEEbiography}[{\includegraphics[width=1in,height=1.25in,clip,keepaspectratio]{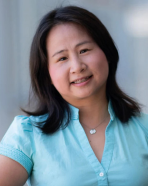}}]
   {Hua Fang} is currently a Professor of the Dept. of Graduate Computer Science and Engineering at Katz School of Science and Health, Yeshiva University. Before that, she was a Professor in the Dept. of Computer and Information Science at the University of Massachusetts Dartmouth. She has been leading Computational Statistics and Data Science Lab which aims at developing computational methods, tools and systems for broader health and behavior studies by integrating and advancing the theories and applications of computational statistics, which is a computational science at the interface of statistics and computer science. She received her PhD degree from the Ohio University. Her current research interests include real-time machine or statistical learning, and visual analytics of wearable biosensor data streams, broadly in e-health and Internet of Things.
\end{IEEEbiography}
    
\end{document}